\documentclass[aps,prc,twocolumn,superscriptaddress,showpacs,nofootinbib,showkeys]{revtex4}

\usepackage{epsfig}
\usepackage{graphicx}
\usepackage{fancyhdr}
\usepackage{pslatex}   
\usepackage{color}
\usepackage{amsmath, amsthm, amssymb} 
\usepackage{natbib}
\usepackage{multirow}
\newcommand{\colormark} {}

\newcommand{\roots}{\mbox{$\sqrt{s_{NN}}$}}

\begin{document}

\title{Constituent quark scaling violation due to baryon number transport}

\author{J.C. Dunlop}
  \affiliation{Brookhaven National Laboratory, Upton New York 11973, USA}
\author{M.A. Lisa}
  \affiliation{Department of Physics, Ohio State University, Columbus, Ohio 43210, USA}
\author{P. Sorensen}
  \affiliation{Brookhaven National Laboratory, Upton New York 11973, USA}

\begin{abstract}
  In ultra-relativistic heavy ion collisions at
  $\roots\approx200$~GeV, the azimuthal emission anisotropy of hadrons
  with low and intermediate transverse momentum ($p_T\lesssim
  4$~GeV/c) displays an intriguing scaling.  In particular, the baryon
  (meson) emission patterns are consistent with a scenario in which a
  bulk medium of flowing quarks coalesces into three-quark (two-quark)
  ``bags.''  While a full understanding of this number of constituent
  quark (NCQ) scaling remains elusive, it is suggestive of a
  thermalized bulk system characterized by colored dynamical degrees
  of freedom-- a quark-gluon plasma (QGP).  In this scenario, one
  expects the scaling to break down as the central energy density is
  reduced below the QGP formation threshold; for this reason,
  NCQ-scaling violation searches are of interest in the energy
  scan program at the Relativistic Heavy Ion Collider (RHIC).
  However, as $\roots$ is reduced, it is not only the initial energy
  density that changes; there is also an increase in the net baryon
  number at midrapidity, as stopping transports entrance-channel
  partons to midrapidity.  \colormark{ This phenomenon can 
  result in violations of simple NCQ scaling.
  Still in the context of the quark coalescence model, we describe a specific pattern for
  the break-down of the scaling that includes
  different flow strengths for particles and their anti-partners.}
  Related complications in the search for recently
  suggested exotic phenomena are also discussed.
\end{abstract}

\maketitle

\section{Introduction}
\label{sec:intro}

\subsection{Quark coalescence in the highest-energy heavy ion collisions}

Hadronization-- the process through which a state characterized by
 colored dynamical partons is resolved into a state of
 colorless hadrons-- is central to the theory of the Strong
 interaction, but remains only incompletely understood.
An important
 aspect of this process has become clear in studies of the forward
 region in hadron-hadron collisions~\cite{Taylor:1975tm} and in high-multiplicity collisions
 of ultra-relativistic heavy ion collisions~\cite{Adler:2003kg,Adams:2003am}: the hadronization of a
 parton is strongly affected by the presence of other partons close in
 phase space.  
Whereas the vacuum hadronization of a single parton
 liberated in a high momentum-transfer ($Q^2$) interaction is described
 in terms of string-breaking scenarios or parameterized in
 fragmentation functions, there is mounting evidence that, in a dense
 phase-space scenario, colored partons essentially ``coalesce'' into
 colorless bound states, much like the formation of light nuclei
 (e.g. deuteron or triton) from free nucleons emitted from a hot zone~\cite{Kapusta:1980zz,Sato:1981ez,Nagle:1996vp}.
Models based on this remarkably simple mechanism, not understood at a
 fundamental level, have enjoyed considerable success at describing the
 ``leading hadron effect''~\cite{Das:1977cp} as well as the multiplicity dependence of
 yields, spectra and momentum anisotropies from heavy ion collisions at
 the highest energies at the Relativistic Heavy Ion Collider (RHIC) ~\cite{Voloshin:2002wa,Hwa:2003bn,Fries:2003vb,Greco:2003xt,Fries:2008hs}.

Even more remarkably, the objects that coalesce appear to be valence
 quarks.  At first, this is surprising, since the dynamical quantities
 in QCD (i.e. the ones that carry momentum) are {\it partons} which
  are overwhelmingly gluons and non-valence quarks.
Indeed, the valence {\it quarks}\footnote{Anti-quarks are treated on
  equal footing with quarks in coalescence models.  Dynamically, we will
  treat antiquarks as simply another variety of quark.}-- three for
  baryons and two for mesons-- were originally invented to explain the
  flavored quantum numbers-- isospin, strangeness, etc-- of the hadrons.
Only later was the connection established between the valence quarks
 and the high momentum-fraction ($x$) fermionic partons.  
Nevertheless,
 the two or three valence quarks represent the lowest Fock states of
 the partonic wavefunction of a hadron, and these appear the relevant
 degrees of freedom;  it is argued that inclusion of higher-order Fock states does not significantly modify the description of the coalescence process and the related phenomenology~\cite{Muller:2005pv}.

The data at RHIC is consistent with
 a partially thermalized system of
 deconfined quarks undergoing collective expansion with an azimuthal
 anisotropy in momentum space proportional to the initial spatial
 anisotropy~\cite{Sorensen:2003wi,Adams:2003am,Sorensen:2009cz}.  
As the system cools, pairs and triplets of neighboring
 quarks coalesce to become the valence quarks (or ``constituent
 quarks'') of mesons and baryons (where the gluons presumably contribute to dressing the valence quarks~\cite{Bhagwat:2007vx,Chang:2010jq}).
The original flow pattern of the
 deconfined quarks leaves a simple and characteristic fingerprint on
 the momentum distribution of the observed hadrons, since a hadron's
 momentum is simply the (vector) sum of the momenta of its valence
 quarks
\begin{equation}
\label{eq:hadronMomentum}
\vec{p}_h = \sum_{i=1}^{n}\vec{p}_{q,i} ,
\end{equation}
where $n=2$ (3) for mesons (baryons).
In the simplest instantaneous $2\rightarrow 1$ or $3\rightarrow 1$ coalescence process, only three
  of the four momentum components are conserved; either energy or momentum conservation is violated~\cite{Fries:2008hs}.  
The most important
 features of quark coalescence that we discuss in this paper are not
 substantially altered in a more complete treatment of the phenomenon,
 accounting for energy and entropy conservation
 effects~\cite{Ravagli:2007xx,Krieg:2007sx,Fries:2008hs}.

In particular, the quarks' radial flow-- the enhancement towards
higher transverse momentum ($p_T$) due to pressure-driven expansion of
the bulk source-- is reflected most strongly in the three-quark
baryons.  
Thus, coalescence provides a natural explanation for the
 ``anomolous baryon enhancement'' at intermediate $p_T$ observed in the
 highest-energy heavy ion collisions at RHIC~\cite{Adler:2003kg}.  
Similarly, the nuclear
 modification factor-- the scaled ratio $R_{AA}\left(p_T\right)$ of
 transverse momentum distributions from heavy ion and $p+p$
 collisions-- shows a clear separation into mesons and baryons~\cite{Adams:2003am,Sorensen:2009cz}.

In heavy ion collisions, the azimuthal anisotropy of the momentum
 distribution is characterized by Fourier components,
\begin{equation}
\label{eq:vnDefinition}
\frac{dN}{d\phi} \propto 1 + 2\sum_{n=1}v_n\left(p_T\right)\cos n\phi ,
\end{equation}
where $\phi$ is measured relative to the direction of the impact
 parameter~\citep[c.f.][for a full discussion]{Voloshin:2008dg}.
Of
 particular interest is the elliptic flow parameter
 $v_2\left(p_T\right)$, which is strongly sensitive to the equation of
 state of QCD matter (e.g. speed of sound) as well as tranport
 coefficients like viscosity.  
For small values of $v_2$ and narrow
 hadronic wavefunctions, the elliptic flow parameters of a bulk system
 of quarks ($a$ and $b$) and the mesons into which they coalesce, are
 related by~\cite{Fries:2008hs}
\begin{equation}
\label{eq:v2quarksMesons}
v_2^M\left(p_T\right) = v_2^a\left(x_ap_T\right) + v_2^b\left(x_bp_T\right) 
\end{equation}
for fixed momentum fractions $x_a$ and $x_b$ ($x_a+x_b=1$), with an
 analogous equation for baryons.

In the event that the constituent quarks ($a$ and $b$) have the same
 elliptic flow before hadronization, Equation~\ref{eq:v2quarksMesons}
 leads to the number-of-constituent-quark (NCQ) scaling pattern 
 observed at RHIC,
\begin{equation}
\label{eq:NCQscaling1}
v_2^h\left(p_T^h\right) = nv_2^q\left(p_T^h/n\right) ,
\end{equation}
where $n=2$ (3) for mesons (baryons).  
In this scenario, all mesons
 should follow one common $v_2\left(p_T\right)$ curve, and all baryons
 another.  
The two should be related via
\begin{equation}
\label{eq:NCQscaling2}
\frac{v_2^B\left(p_T/3\right)}{3} = 
\frac{v_2^M\left(p_T/2\right)}{2} \quad
\left(= v_2^q\left(p_T\right)\right) .
\end{equation}

\subsection{Violation of NCQ scaling}

The observed satisfaction of Eq.~\ref{eq:NCQscaling2} in Au+Au
 collisions at RHIC was one of the most compelling indications that
 deconfined partonic degrees of freedom were playing a dynamic role in
 the bulk medium of the early phase.  
Consequently, the {\it breakdown}
 of this scaling as the beam energy is reduced, is an
 important signal in the RHIC energy scan program~\cite{Aggarwal:2010cw}.  
It may
 indicate that the initial energy density of the system is below the
 threshold for QGP creation, pinpointing the phase transition between
 confined and deconfined QCD matter.

Additionally, it has been recently proposed~\cite{Burnier:2011bf} that a deconfined
  bulk system at finite baryon density may acquire an electric quadrupole
  moment due to chiral magnetic waves in the plasma.
This interesting phenomenon would break the degeneracy between $v_2$ for
  positive and negative pions, also clearly breaking NCQ scaling.

It is important to consider other \colormark{ less exotic}
 mechanisms that may also cause violation of the scaling represented by
 Eq.~\ref{eq:NCQscaling2}.
In particular, we recall that this
 scaling should hold if all quarks (and antiquarks) have the same
 underlying flow distribution ($v_2^q$ in
 equation~\ref{eq:NCQscaling2}). This would be a natural conseqence of thermalization.
The scenario where all constituent quarks have the same $v_2^q$ we call NCQ$_1$.


\colormark{
In this paper, we examine whether the breakdown of Eq.~\ref{eq:NCQscaling2} would necessarily signal
  that the hadrons are not resulting from the coalescence of flowing constituent quarks.
In particular, we discuss a minimal extension of the NCQ$_1$ model that retains
  constituent quarks as the dynamical degrees of freedom and coalescence as the
  hadronization mechanism.
However, the assumption that all quarks have the same $v_2^q$ is discarded due to the well-recognized phenomenon of
  baryon stopping, which is increasingly important at lower energies.
In particular, we recall that the transport of baryon number from the entrance channel
  to midrapidity (``stopping'') is increasingly important at lower energies.
Since we continue to work in the dynamical constituent quark paradigm,  
  this means that $u$ and $d$ quarks
  transported from $y=y_{\rm beam}$ to $y=0$ surely suffer multiple collisions with each other.
Meanwhile, at the lower energies in question, the quark-antiquark pairs created from the vacuum
  may experience relatively fewer.
} In a picture where $v_2^q$ is developed through collisions, it is not unreasonable to expect that quarks transported from $y=y_{\rm beam}$ to $y=0$ will develop a larger $v_2$.

\colormark{
Such considerations will lead to a specific pattern for the breakdown of Eq.~\ref{eq:NCQscaling2}.
Without invoking exotic phenomena, this simple scenario also implies that the degeneracy of $v_2$
  for particles and their anti-partners will be broken in a specific way.
Not only its prediction for $v_2\left[\pi^+\right]-v_2\left[\pi^-\right]$,
  but also that for $v_2\left[K^+\right]-v_2\left[K^-\right]$, can be compared to data and predictions
  from more complicated models.
}


In the following section, we briefly review the energy dependence of
  stopping in heavy ion
  collisions-- transport of baryon number from the high-rapidity region
  of the entrance channel to midrapidity in the exit channel.  
\colormark{
We also use measured particle yields to estimate the fraction of $u$ and $d$ quarks at midrapidity that
  would arise from baryon stopping at two collision energies.
}
\colormark{
In section~\ref{sec:TwoComponent}, we consider quantitatively a two component model for quark number scaling NCQ$_2$,
 in which the phenomenon of stopping leads to at least two samples of quarks with different $v_2^q$ values which then coalesce into hadrons.
%
%
%
For tractability, we idealize this non-thermal distribution in a two-component formalism: transported
 quarks follow one flow profile and produced quarks another.
}
We briefly summarize in section~\ref{sec:conclusions}.

\section{Transport of entrance-channel quarks to midrapidity}
\label{sec:stopping}

\colormark{
In this section, we briefly review the phenomenon of baryon stopping in heavy ion collisions.
We then use hadron yields measured by the NA49/SPS collaboration
  to estimate the
  fraction of $u$ and $d$ quarks at midrapidity that would arise from stopping.
These fractions will be used in section~\ref{sec:TwoComponent} as input to a simple model
  to predict the breakdown of NCQ scaling, Eq.~\ref{eq:NCQscaling2}.
}

\subsection{Stopping in heavy ion collisions}
In the RHIC energy scan program, the beam energy is varied to
  modify the initial conditions of the hot QCD system created.
In
  addition to changing the energy density of the initial state, it is
  well-accepted that, due to baryon stopping, the baryo-chemical potential $\mu_B$ of the system
  is larger at lower $\roots$.

Baryon stopping, the transport of baryon number from its initial
  location at beam rapidity towards the initially baryon-free region at
  mid-rapidity, is most directly measured via the rapidity distribution
  of net protons (the number of protons minus antiprotons).
At $\roots\approx$ 5~GeV, the rapidity distribution is peaked at midrapidity.
As the collision energy is increased, the distribution peaks at increasingly forward rapidity.
This behavior has been
  parametrized as an average rapidity loss, which increases from
  approximately 1 unit at $\roots\approx$ 5 GeV to 1.7 unit at
  $\roots\approx$ 17 GeV, with a smaller rise towards 2 units by the
  highest RHIC $\roots$ of 200 GeV\cite{Arsene:2009fn}.  
This rise in
  rapidity loss is less rapid than the rise in the beam rapidity with
  increasing $\roots$, leading to a decreasing population of net baryon
  number at mid-rapidity.

More detailed statistical model calculations, based on measurements of
  the yields of a range of particles, agree with this general
  conclusion.  
Within these models, with increasing $\roots$ the net
  baryon density first rises, achieves a maximum at $\roots\approx$ 8
  GeV, and then falls for higher \roots~\cite{Randrup:2006uz}.  
The
  system transitions from one with entropy density dominated by baryons
  at low energy to one dominated by mesons at high energy, with equal
  fractions at $\roots\approx$ 8 GeV~\cite{Oeschler:2009im}.  The 
  {\it fractional} importance of transported quarks to the system's
  evolution grows rapidly with decreasing $\roots$ in the lower end of
  the region probed by the RHIC Beam Energy scan, from $\roots\approx$
  15 GeV to 7.7 GeV.

\subsection{Estimates for the stopping contribution to light quark yields at midrapidity}

\colormark{
In section~\ref{sec:TwoComponent}, we develop a simple model in which transported $u$ and $d$ quarks
  have stronger flow than do produced quarks (including produced $u$ and $d$ quarks).
An important ingredient to this model is the fraction of $u$ ($d$) quarks present at midrapidity
  that arise from baryon number transport.
In particular, we want the fraction
\begin{equation}
\label{eq:X_u^T}
X_{u^T}\equiv\frac{N_{u^T}}{N_{u^T}+N_{u^P}} ,
\end{equation}
where $N_{u^T}$ is the number of $u$ quarks from the incoming heavy ions transported to midrapidity,
  and $N_{u^P}$ is the number of $u$ quarks produced from $u-\bar{u}$ pair production at midrapidity.
The fraction $X_{d^T}$ is defined similarly.
}

\colormark{
To estimate $X_{u^T}$ and $X_{d^T}$, we use measured midrapidity yields of common particles from central Pb+Pb collisions by the NA49/SPS
  Collaboration at $\sqrt{s_{NN}}=6.41$ and~8.86~GeV.
Tables~\ref{tab:6.41Xestimates} and~\ref{tab:8.86Xestimates} list the measured yields of hadrons and their
  constituent quarks.
}

\colormark{
Since produced quarks and antiquarks are formed in pairs, the number of transported $u$ quarks
  is given by the imbalance between the total number of $u$ and $\bar{u}$ quarks; 
$X_{u^T}=\left(N_{u}-N_{\bar{u}}\right)/N_{u}$.
Table~\ref{tab:Xvalues} lists the fractions for the two energies.
}

Several comments are in order.
Firstly, at these energies, roughly half of the light constituent quarks at midrapidity
  orginate from the colliding nuclei; clearly stopping cannot be ignored.
Secondly, the fraction of $d$ quarks transported from the $y=y_{\rm beam}$ is greater
  than the fraction of $u$ quarks, simply as a consequence of the isospin of the entrance channel.
Thirdly, the imbalance in $s$ and $\bar{s}$ quarks in tables~\ref{tab:6.41Xestimates} and~\ref{tab:8.86Xestimates}
  reminds us that our estimates are just that.

\begin{table}[t!]
\begin{tabular}{|l|l|l|l|l|l|l|l|}
\hline
hadron          & yield & $u$   & $d$   & $s$   & $\bar{u}$ & $\bar{d}$ & $\bar{s}$ \\
\hline
$\pi^+$         & 72.9  & 72.9  & ~     & ~     & ~         & 72.9      & ~         \\
$\pi^-$         & 84.8  & ~     & 84.8  & ~     & 84.8      & ~         & ~         \\
$\pi^0$ (*)     & 78.85 & 39.43 & 39.43 & ~     & 39.43     & 39.43     & ~         \\
$\phi$          & 1.17  & ~     & ~     & 1.17  & ~         & ~         & 1.17      \\
K$^+$           & 16.4  & 16.4  & ~     & ~     & ~         & ~         & 16.4      \\
K$^-$           & 5.58  & ~     & ~     & 5.58  & 5.58      & ~         & ~         \\
K$^0$ (*)       & 10.99 & ~     & 10.99 & ~     & ~         & ~         & 13.84     \\
$\bar{\rm K}^0$ (*) & 10.99 & ~     & ~     & 10.99 & ~         & 10.99     & ~         \\
p               & 46.1  & 92.2  & 46.1  & ~     & ~         & ~         & ~         \\
n (*)           & 70.84 & 70.84 & 141.68& ~     & ~         & ~         & ~         \\
$\Lambda$       & 13.4  & 13.4  & 13.4  & 13.4  & ~         & ~         & ~         \\
$\Xi^-$         & 0.93  & ~     & 0.93  & 1.86  & ~         & ~         & ~         \\
$\bar{\rm p}$   & 0.06  & ~     & ~     & ~     & 0.12      & 0.06      & ~         \\
$\bar{\rm n}$ (*)  & 0.04  & ~     & ~     & ~     & 0.04      & 0.08      & ~         \\
$\bar{\Lambda}$ & 0.1   & ~     & ~     & ~     & 0.1       & 0.1       & 0.1       \\
\hline
Sum             & ~     & 305.17 & 337.33 & 33    & 130.07    & 123.56    & 28.66     \\
\hline
\end{tabular}
\caption{
\label{tab:6.41Xestimates}
Left two columns: midrapidity yields of common particles from central Pb+Pb collisions measured by the NA49/SPS
  Collaboration~\cite{Afanasiev:2002mx,Anticic:2004yj,Alt:2006dk,Alt:2007fe,Alt:2008qm,Alt:2008iv,PhysRevC.83.014901,NA49dataList}
 at $\sqrt{s_{NN}}=6.41$~GeV.
Starred hadrons are not measured, but estimated from other hadrons.
In particular, $dN\left[\pi^0\right]/dy=0.5\left(dN\left[\pi^+\right]/dy+dN\left[\pi^-\right]/dy\right)$,
  $dN\left[{\rm K}^0\right]/dy=dN\left[\bar{{\rm K}}^0\right]/dy=0.5\left(dN\left[{\rm K}^+\right]/dy+dN\left[{\rm K}^-\right]/dy\right)$,
  $dN\left[{\rm n}\right]=1.54dN\left[{\rm p}\right]$, and 
  $dN\left[\bar{\rm n}\right]=1.54^{-1}dN\left[\bar{\rm p}\right]$.
The factor 1.54 is the neutron-to-proton ratio of Pb.
Right six columns: midrapidity yield of constituent quarks for each hadron.
}
\end{table}

\begin{table}[t!]
\begin{tabular}{|l|l|l|l|l|l|l|l|}
\hline
hadron          & yield & $u$   & $d$   & $s$   & $\bar{u}$ & $\bar{d}$ & $\bar{s}$ \\
\hline
$\pi^+$         & 96.6  & 96.6  & ~     & ~     & ~         & 96.6      & ~         \\
$\pi^-$         & 106.1 & ~     & 106.1 & ~     & 106.1     & ~         & ~         \\
$\pi^0$ (*)     & 101.35& 50.68 & 50.68 & ~     & 50.68     & 50.68     & ~         \\
$\phi$          & 1.16  & ~     & ~     & 1.16  & ~         & ~         & 1.16      \\
K$^+$           & 20.1  & 20.1  & ~     & ~     & ~         & ~         & 20.1      \\
K$^-$           & 7.58  & ~     & ~     & 7.58  & 7.58      & ~         & ~         \\
K$^0$ (*)       & 13.84 & ~     & 13.84 & ~     & ~         & ~         & 13.84     \\
$\bar{\rm K}^0$ (*) & 13.84 & ~     & ~     & 13.84 & ~         & 13.84     & ~         \\
p               & 41.3  & 82.6  & 41.3  & ~     & ~         & ~         & ~         \\
n (*)           & 63.5  & 63.5  & 127   & ~     & ~         & ~         & ~         \\
d               & 1.02  & 3.06  & 3.06  & 3.06  & ~         & ~         & ~         \\
$\Lambda$       & 14.6  & 14.6  & 14.6  & 14.6  & ~         & ~         & ~         \\
$\Xi^-$         & 1.15  & ~     & 1.15  & 2.3   & ~         & ~         & ~         \\
$\bar{\rm p}$   & 0.32  & ~     & ~     & ~     & 0.64      & 0.32      & ~         \\
$\bar{\rm n}$ (*)  & 0.21  & ~     & ~     & ~     & 0.21      & 0.42      & ~         \\
$\bar{\Lambda}$ & 0.33  & ~     & ~     & ~     & 0.33      & 0.33      & 0.33      \\
$\bar{\Xi}^+$   & 0.07  & ~     & ~     & ~     & ~         & 0.07      & 0.14      \\
\hline
Sum             & ~     & 331.14 & 357.73 & 39.48 & 165.54    & 162.26    & 35.57     \\
\hline
\end{tabular}
\caption{
\label{tab:8.86Xestimates}
The same as for table~\ref{tab:6.41Xestimates}, but for $\sqrt{s_{NN}}=8.86$~GeV collisions.
}
\end{table}

\begin{table}[ht]
\begin{tabular}{|l|l|l|}
\hline
$\sqrt{s_{NN}}$  & $X_{u^T}$ & $X_{d^T}$ \\
\hline
6.41 GeV         & 0.57      & 0.63      \\
8.86 GeV         & 0.50      & 0.55      \\
\hline
\end{tabular}
\caption{
\label{tab:Xvalues}
Based on the data in tables~\ref{tab:6.41Xestimates} and~\ref{tab:8.86Xestimates},
  the fraction of $u$ and $d$ quarks at midrapidity, that originate from stopping of quarks in the colliding Pb nuclei.
See text for details.
}
\end{table}

\section{Violations of simple NCQ scaling in a two-component scenario}
\label{sec:TwoComponent}

\colormark{
As discussed in section~\ref{sec:intro}, the violation of NCQ scaling, eq.~\ref{eq:NCQscaling2} and the breakdown of
  the degeneracy of $v_2$ between particles and antiparticles, implies a lack of kinetic thermalization
  of the dynamic system.
This may arise if the energy of the collision falls below the threshold to produce a flowing system of deconfined
  partons, so that particle-specific hadronic cross-sections determine each hadron's flow strength.
Alternatively, it may arise from more interesting phenomena such as chiral magnetic waves.
Our approach in this paper is a minimalist one, asking whether such effects may be expected without
  invoking exotic phenomena or abandoning a scenario of flowing quarks which coalesce into flowing hadrons.
}

\colormark{
The absence of complete kinetic thermalization at low energies ($\sqrt{s_{NN}}\lesssim 30$~GeV) has long
  been recognized, based on the severe failure of hydrodynamics to reproduce hadronic $v_2$~\cite{Alt:2003ab}.
In short, at these energies, the constituents of the system do not rescatter sufficiently to achieve
  thermalization.
As we have discussed, at these energies baryon transport from the entrance channel plays a huge role.
We wish to test the robustness of the constituent quark paradigm, so the transported baryon number is
  represented by transported $u$ and $d$ quarks.
The very fact that they have been transported over a significant rapidity range attests to the likelihood
  that these quarks, in any event, have suffered many scatterings.
We make the plausible postulate that transported quarks experience more scatterings than produced ones
  at these energies, hence approaching the thermal limit more closely and developing a larger $v_2$.
}

\colormark{
Clearly, the resulting non-thermal quark momentum distribution reflects a continuum of quarks
  rescattering more or less before coalescence.
In order to render the problem tractable, we model the situation in a simple limit of two populations:
  transported quarks and produced quarks, with the former population characterized by a stronger flow
  than the latter.
We emphasize that this is a simplification in order to make a point: we do not propose that there are
  really two distinct thermalized fluids created in a heavy ion collision.
}

\colormark{
Hence, we have two populations of constituent quarks with distinct flow
   fields, $v_2^{q^P}$ and $v_2^{q^T}$,
   for produced ($u^{P}$, $\bar{u}^{P}$, $d^{P}$, $\bar{d}^{P}$, $s^{P}$, $\bar{s}^{P}$) 
  and transported ($u^{T}$, $d^{T}$) quarks, respectively.
In this simplest two-component
  model, a hadron's elliptic flow parameter is given by
\begin{equation}
\label{eq:TlowFields}
v_2^h\left(p_T\right) =       
     \sum_{i=1}^n\left[ X_{q_i^T}v_2^{q_i^T}\left(p_T/n\right) + \left(1-X_{q_i^T}\right)v_2^{q_i^P}\left(p_T/n\right)\right] ,
\end{equation}
where $X_{q_i^T}$ is the fraction of quark species $q_i$ that originates from baryon stopping,
  as discussed in section~\ref{sec:stopping}.
As per the discussion in that section, reasonable estimates are $X_u=0.50$ and $X_d=0.55$.
Naturally, $X_{\bar{u}^T}=X_{\bar{d}^T}=X_{s^T}=X_{\bar{s}^T}=0$.
}

\begin{figure}[]
{\centerline{\includegraphics[width=0.5\textwidth]{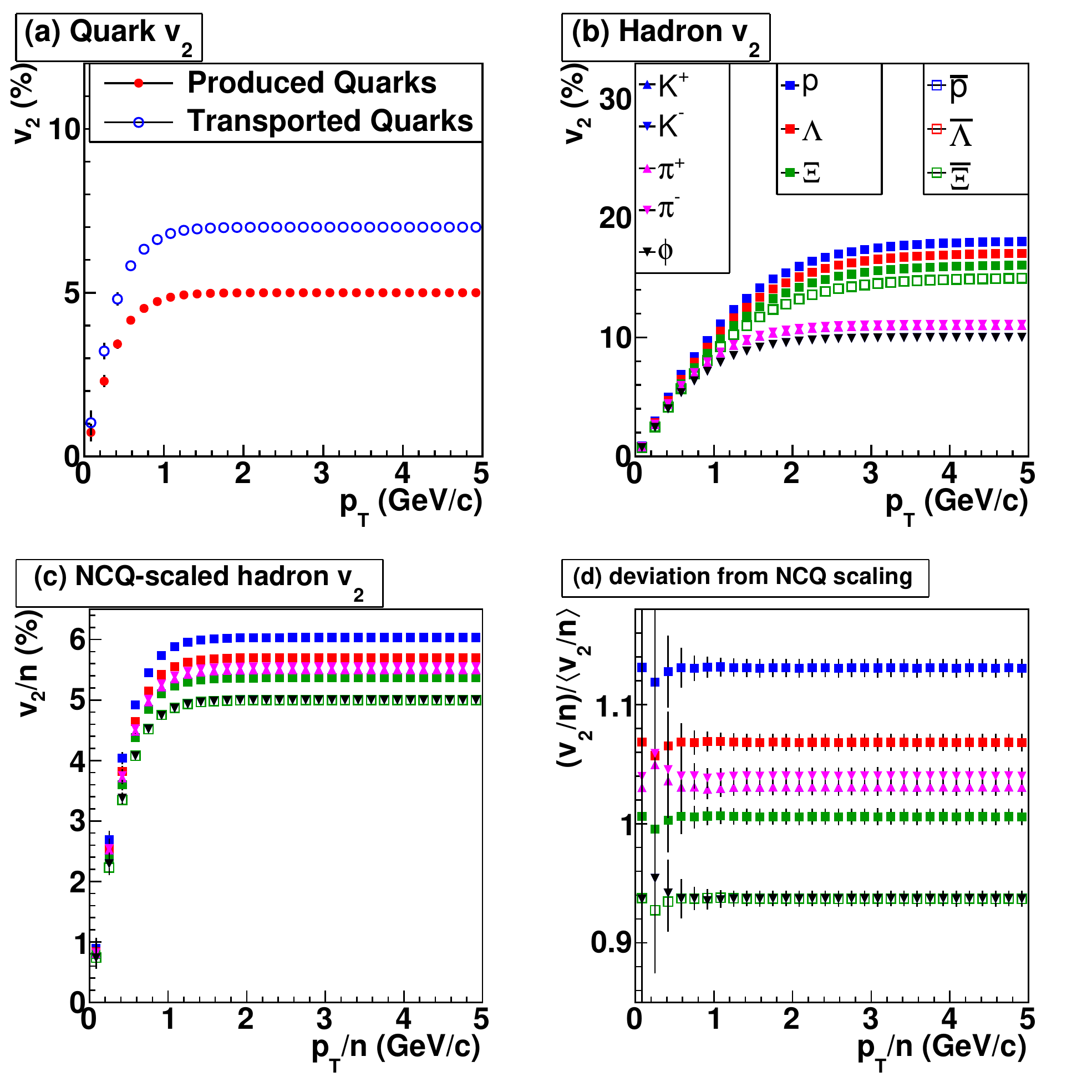}}}
\caption{
(Color online)
NCQ$_2$, the simplest generalization of the NCQ$_1$ model, in which the transported up and down quarks
  have a 40\% stronger intrinsic $v_2$ than
  do the produced quarks, which themselves all have the same $v_2$.
The fraction of $u$ and $d$ quarks that are transported is 50\% and 55\%, respectively.
See text for details.\\
Panel (a): Intrinsic $v_2$ of all quarks.\\
Panel (b): Hadron $v_2$ based on coalescence of quarks shown to the left.\\
Panel (c):  NCQ-scaled hadron flow, $v_2\left(p_T/n\right)/n$.\\
Panel (d):  Deviation from simple NCQ scaling-- the curves from the lower left panel, divided by their average.\\
In panels (b)-(d), ${\rm K}^+$ data points lie beneath those for $\pi^+$, ${\rm K}^-$ 
  data points lie beneath those of $\phi$, and the data points for all anti-baryons are coincident.
\label{fig:AllProducedTheSame}
}
\end{figure}

Figure~\ref{fig:AllProducedTheSame} shows a example of the resulting $v_2$ from our simple NCQ$_2$ scenario.
For the purpose of illustration, for these calculations, we had to assume some functional form for
  the quark elliptic flow.
We chose the same functional form for both produced and transported quarks:
\begin{equation}
\label{eq:v2FunctionalForm}
v_2\left(p_T\right)=M\tanh\left(p_T/\left(0.5~{\rm GeV/c}\right)\right) .
\end{equation}
For the example in figure~\ref{fig:AllProducedTheSame}, $M=0.07$ for transported
  quarks and $M=0.05$ for produced quarks.
The choice of this particular functional form is rather arbitrary and does not affect
  the points we make below.

Clearly, simple NCQ scaling (equation~\ref{eq:NCQscaling2}) is violated, with an unavoidable species-dependent signature.
In particular, one finds:
\begin{align}
\label{eq:BreakdownPattern}
v_2\left[\pi^-=d\bar{u}\right] &> v_2\left[\pi^+=u\bar{d}\right] \nonumber \\
v_2\left[{\rm K}^+=u\bar{s}\right] &> v_2\left[{\rm K}^-=\bar{u}s\right] \nonumber \\
v_2\left[p=uud\right] &> v_2\left[\bar{p}=\bar{u}\bar{u}\bar{d}\right] \\
v_2\left[\Lambda=uds\right] &> v_2\left[\bar{\Lambda}=\bar{u}\bar{d}\bar{s}\right] \nonumber\\
v_2\left[p=uud\right] &> v_2\left[\Lambda=uds\right] \nonumber \\
\left( v_2\left[p=uud\right] - v_2\left[\bar{p}=\bar{u}\bar{u}\bar{d}\right] \right) &>
  \left(v_2\left[\Lambda=uds\right] - v_2\left[\bar{\Lambda}=\bar{u}\bar{d}\bar{s}\right]\right) . \nonumber
\end{align}

It is interesting that the ordering of $v_2$ for positive and negative pions is the same as
  that predicted due to chiral magnetic wave effects~\cite{Burnier:2011bf}.
We also find that the charge-ordering for kaons ($v_2\left[{\rm K}^+\right]>v_2\left[{\rm K}^-\right]$)
  is opposite to that for pions.
The chiral magnetic wave effect would generate the {\it same} charge-ordering for pions as for kaons,
  thus providing a testable distinction between the chiral magnetic model and our stopping-based model.
However, hadronic effects (e.g. the smaller cross-section for ${\rm K}^-$) may complicate the interpretation
  of such a test~\cite{Dima:PC}.

\begin{figure}[]
{\centerline{\includegraphics[width=0.5\textwidth]{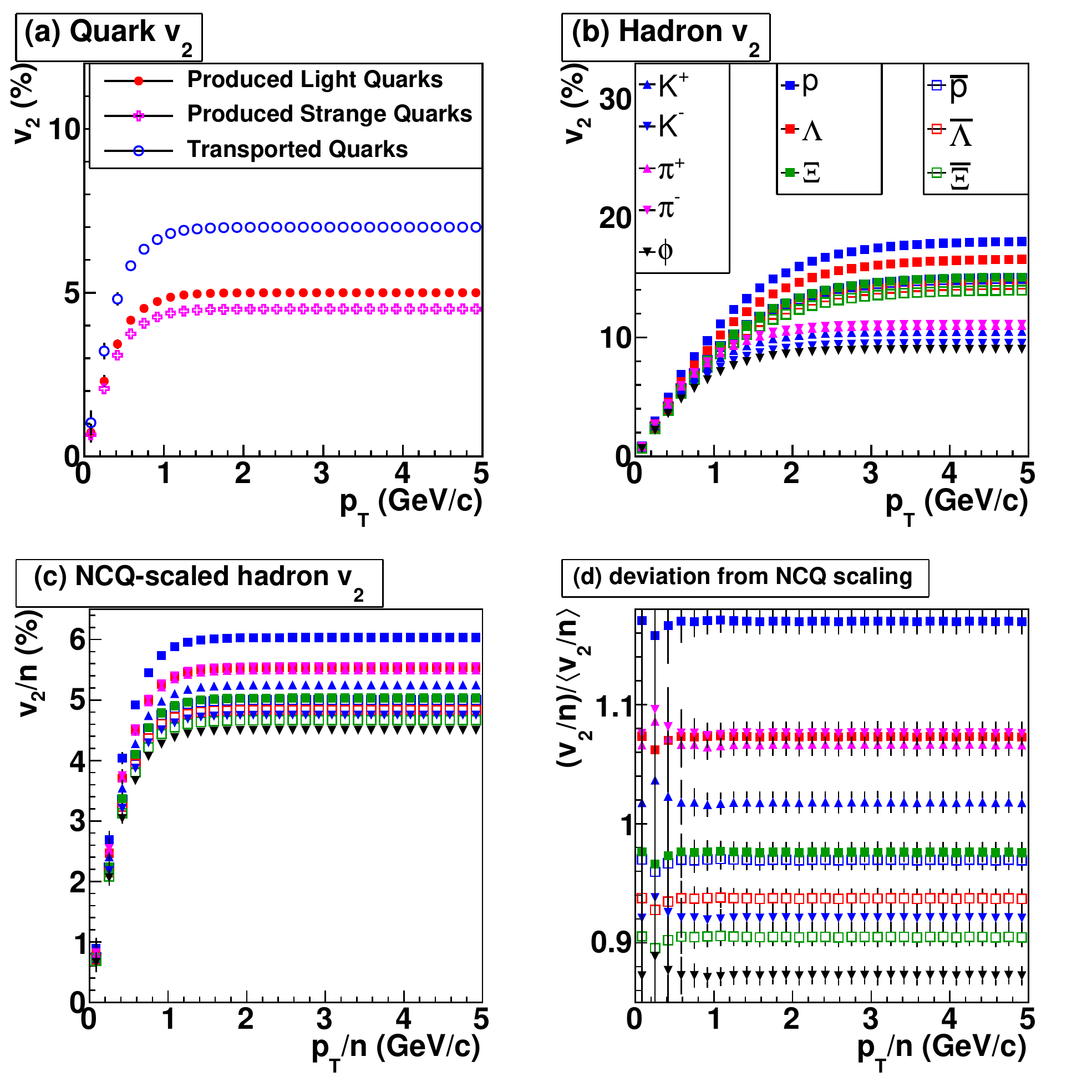}}}
\caption{
(Color online)
The same as in figure~\ref{fig:AllProducedTheSame}, but the produced strange quarks
  have 10\% less intrinsic $v_2$ than do the produced light quarks.
See text for details.
\label{fig:StrangeProducedQuarksFlowLess}
}
\end{figure}

Comparing the anisotropies of particles and their anti-partners, as listed in equation~\ref{eq:BreakdownPattern},
  is straightforward and relatively unambiguous.
The details of cross-species comparisons can depend more on the particular functional forms used for the quark
  flow profiles, the weighting factors $X_{q_i^T}$ and whether the latter depend on $p_T$.
For the simple case we have considered, $\frac{v_2\left[p\right]}{3}>\frac{v_2\left[\pi^\pm\right]}{2}$;
  the entire species dependences can be seen in panels (d) of figures~\ref{fig:AllProducedTheSame} and~\ref{fig:StrangeProducedQuarksFlowLess}.

Our primary points have been made already in this simple model, but we mention an additional complication.
If there is insufficient rescattering to fully thermalize the light produced quarks, then the heavier
  strange produced quarks are likely to be even less thermalized.
In this case, $v_2^{s^P}<v_2^{u^P,d^P}$; similar considerations have been discussed by Lin and
  Ko~\cite{Lin:2002rw}.
Figure~\ref{fig:StrangeProducedQuarksFlowLess} shows the situation when the functional form of
  equation~\ref{eq:v2FunctionalForm} describes the flow of all quarks, as before, but now
  M=0.045 for the strange quarks.
In this case, the degeneracies 
  (e.g. $v_2\left[{\rm K}^-\right]=v_2\left[\phi\right]$) 
  seen in figure~\ref{fig:AllProducedTheSame} and listed in its caption are broken; all hadrons have
  distinct elliptic flow curves.

Additional reasonable complications can be considered.
Clearly, the functional forms used for the quark flow can be varied from the simple form (eq.~\ref{eq:v2FunctionalForm})
  used here.
Furthermore, one may reasonably argue that the fraction of light quarks arising from transport ($X_{q_i^T}$) should
  depend on $p_T$; we have treated it as a constant for simplicity.
Exploring such considerations amounts to tuning the model.
We leave such explorations for later comparison and fitting when data become available.

\section{Discussion and summary}
\label{sec:conclusions}

The success of NCQ scaling of elliptic flow at \roots=200~GeV has been
  one of the most striking observations at RHIC, strongly suggesting the
  creation of a flowing, thermalized bulk system of quarks that coalesce into hadrons.
Hence, observing the violation of this scaling as
  \roots~ is decreased could be of crucial importance, both for
  validating the simple dynamical constituent quark model, and for pinpointing the conditions required to undergo the
  deconfinement phase transition.  
Furthermore, recent theoretical
  predictions suggest that a chiral magnetic wave effect may reveal itself by
  inducing a different flow for positive and negative pions~\cite{Burnier:2011bf}.
The observation of NCQ scaling violations would thus be
  potentially far-reaching.

It is important, therefore, to explore less exotic reasons for any
  scaling violations.  
We have discussed one simple scenario here, which
  requires neither a fundamental difference in the phase of QCD matter
  in the measured energy range, nor a new exotic effect.

The model predicts an unavoidable species-dependent pattern for the breakdown of NCQ scaling
  and depends on only two assumptions.
Firstly, it assumes that, just as at top RHIC energies, the system can be described in terms
  of constituent quarks that coalesce into hadrons as the system cools.
Secondly, it assumes that quarks transported from beam rapidity to midrapidity suffer more
  violent scatterings than do quarks produced at midrapidity at low $\sqrt{s_{NN}}$.
We {\it simplified} the situation by treating the system as two distinct quark populations, but our
  main points do not depend on this simplification.

(Baryon transport from the entrance channel is another important ingredient of the model, but
  its relevance is far from an assumption; the phenomenon of stopping is well known and the
  isospin effect ($X_{d^T}>X_{u^T}$) is obvious
  and based on data, as discussed in section~\ref{sec:stopping}.)

The second of our two assumptions seems at least very plausible.
It is clear that at low energies, the system does not have sufficient density or energy to fully thermalize-- the
  dynamical constituents do not scatter enough.
Unlike the produced particles born at midrapidity, however, the transported quarks had to undergo
  several collisions just to reach midrapidity, after which they could rescatter further.

It is the first assumption-- that even at low energies where scaling violations might be found, the system
  is well-described by a flowing system of constituent quarks-- that seems most questionable.
Nevertheless, our task has been to explore the implications of its validity even at low $\sqrt{s_{NN}}$.
We have found an unambiguous species dependendence of $v_2$ listed in equations~\ref{eq:BreakdownPattern}.
Quantitative details depend on tuning which we do not consider in this first study.
Detailed comparisons with experimental data should be performed, but
  we have shown that violation of NCQ scaling or particle-antiparticle $v_2$ degeneracy themselves is insufficient
  to claim either the crossing of the deconfinement threshold or exotic phenomena.

\section*{Acknowledgements}
We would like to thank Dr. Michael Mitrovski and Dr. Rosi Reed for helpful discussions.
This work supported by the U.S. National Science Foundation under Grant PHY-0970048 and by
  the Offices of NP and HEP within the U.S. DOE Office of Science under the contracts
  DE-FG02-88ER40412 and DE-AC02-98CH10886.


\end{document}